\documentclass[preprint,preprintnumbers,amsmath,amssymb]{revtex4}

\usepackage{graphicx}
\usepackage{dcolumn}
\usepackage{bm}

\begin{document}

\preprint{Nuclei in the Cosmos IX, June 25-30 2006, CERN, Geneva, Switzerland}

\title{First measurements of the total and partial stellar cross section to the $s$-process branching-point $^{79}$Se}

\author{I. Dillmann}\email{iris.dillmann@ik.fzk.de}
 \altaffiliation[also at ]{Departement f\"ur Physik und Astronomie, Universit\"at Basel.}
\author{M. Heil}
\author{F. K\"appeler}
 \affiliation{Institut f\"ur Kernphysik, Forschungszentrum Karlsruhe, Postfach 3640, D-76021 Karlsruhe, Germany}

\author{T. Faestermann}
\author{K. Knie}
\author{G. Korschinek}
\author{M. Poutivtsev}
\author{G. Rugel}
\affiliation{Fakult\"at f\"ur Physik, Technische Universit\"at M\"unchen, D-85747 Garching}

\author{A. Wallner}
 \affiliation{Vienna Environmental Research Accelerator, Institut f\"ur Isotopenforschung und Kernphysik, Universit\"at Wien, A-1090 Wien}

\author{T. Rauscher}
   \affiliation{Departement Physik und Astronomie, Universit\"at Basel, Klingelbergstrasse 82, CH-4056 Basel, Switzerland}

\begin{abstract}
Although $^{79}$Se represents an important branching in
the weak $s$ process, the stellar neutron capture cross sections
to this isotope have not yet been measured experimentally. In this
case, experimental data is essential for evaluating the important
branching in the $s$-process reaction path at $^{79}$Se. The total
cross section of $^{78}$Se at a stellar energy of $kT$ = 25 keV
has been investigated with a combination of the activation
technique and accelerator mass spectrometry (AMS), since off\-line
decay counting is prohibitive due to the long terrestrial half
life of $^{79}$Se (2.80$\pm$0.36 $\times$10$^5$~y \cite{HJJ02}) as
well as the absence of suitable $\gamma$-ray transitions. The
preliminary result for the total Maxwellian averaged cross section
is $<\sigma>_{30~keV}$= 60.1$\pm$9.6~mbarn, signif\-icantly lower
than the previous \mbox{recommended} value. In a second
measurement, also the partial cross section to the 3.92~min-isomer
was determined via $\gamma$-spectroscopy and yielded
$<\sigma>_{30~keV}$(part.)= 42.0$\pm$2.0~mbarn.
\end{abstract}

\maketitle

\section{Introduction}
The origin of the elements heavier than iron can be almost
completely ascribed to neutron capture processes characterized by
much longer and much shorter time scales compared to average
$\beta$-decay half-lives. The respective nucleosynthesis
processes, known as the slow ($s$) and the rapid ($r$) neutron
capture process, contribute in approximately equal parts to the
total elemental abundances in the mass range above iron. The $s$
process can be subdivided into two fractions, corresponding to
different mass regions, temperature ranges and neutron exposures.
The "weak" component, responsible for the elements with A$<$90, is
driven by the neutron production via the
$^{22}$Ne$(\alpha,n)$$^{25}$Mg reaction at temperatures of T= 2-3
$\times$10$^8$ K. The "main" component is mainly based on the
$^{13}$C$(\alpha,n)$$^{16}$O reaction, which operates at
temperatures of 10$^8$ K. In this case the
$^{22}$Ne$(\alpha,n)$$^{25}$Mg contributes only a small part of
the total exposure, but affects the f\-inal abundance pattern.
This main component is producing the $s$ abundances in the region
A$\leq$90$\leq$A$\leq$208. The astrophysical environment for the
weak $s$ process are massive stars with core He and shell C
burning, whereas the main component needs He shell f\-lashes in
low mass TP-AGB stars.

\begin{figure*}[!hb]
\includegraphics{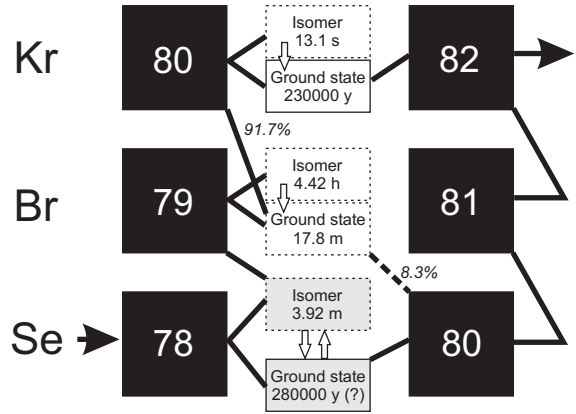}
\caption{\label{branch}Sketch of the $s$-process f\-low in the
Se-Br-Kr region. The half-lives given here are terrestrial
half-lives.}
\end{figure*}

Our measurements refer to the region of the "weak" $s$ process.
Among the nuclei involved, the long-lived radioactive isotopes
$^{63}$Ni, $^{79}$Se, and $^{85}$Kr assume key positions, because
the $\beta$-decay rate becomes comparable to the neutron capture
rate ($\lambda_\beta \approx \lambda_n$). This competition leads
to branchings of the synthesis path, as sketched in
Fig.~\ref{branch}. The branching point isotopes are important,
since they can be used for a diagnosis of the temperature and
neutron density by comparing the ratios of isotopes at masses
above and below the respective branching with observations.

The split of the reaction path at $^{79}$Se causes part of the
f\-low to bypass $^{80}$Kr, whereas at $^{82}$Kr both f\-lows
merge back. The strength of the resulting branching is ref\-lected
in the abundance of the $s$-only isotopes $^{80,82}$Kr, which are
shielded from the decay chains of the $r$ process by $^{80,82}$Se.
Analysis of the local abundance pattern in the mass region
80$\leq$A$\leq$82 yields the effective half-life of $^{79}$Se at
the $s$-process site.

It was noticed that the effective branching ratios for $^{79}$Se
and $^{85}$Kr are almost equal. This leads to a conf\-lict, since
the terrestrial half-life of both isotopes differ by roughly four
orders of magnitude. The solution of this puzzle is due to $\beta$
decays from thermally populated excited states. The ground state
$\beta$-decay of $^{79}$Se is f\-irst-forbidden unique, but
allowed decays are possible from the isomer at 95.7~keV, and from
states at 128~keV and 137~keV, which are populated to 1.0\%, 0.4\%
and 1.2\%, respectively, at $kT$=30~keV \cite{WaB82,KBW89}. Apart
from this, stellar decay rates are also inf\-luenced by the high
degree of ionization in the hot stellar plasma. This allows
"bound-state" $\beta$-decays (Q$_\beta$$<$300~keV), which
contribute $\approx$25\% to the stellar rate \cite{TaY83}. It was
shown for $^{79}$Se (see Fig.~7 in \cite{KlK88}), that the
$\beta$-decay rate is almost constant for T$<$10$^8$K, but drops
quickly to the order of a few years when low-lying f\-irst excited
states become thermally populated. Since the temperature
dependence of the half-life is well known from calculations (see
\cite{KlK88} and \cite{TaY83}), the branching at $^{79}$Se can be
interpreted as an $s$-process thermometer \cite{WBK86}.
Nevertheless, this leads to the uncommon fact that the temperature
dependence is better known than the terrestrial half-life.

For $^{78}$Se, no experimental information on its stellar
(n,$\gamma$) cross section to the ground and isomeric state in
$^{79}$Se existed so far. Both cross sections were determined by
means of the activation technique. In combination with AMS this
technique can be extended to hitherto inaccessible cases, e.g. to
reactions producing very long-lived nuclei with very weak or
completely missing $\gamma$ transitions. The application of AMS
counting in stellar neutron reactions, which has been demonstrated
recently for $^{62}$Ni \cite{Nas04} and also $^{58}$Ni
\cite{RDF06}, has the further advantage of being independent of
uncertain $\gamma$-ray intensities or half-lives.

\section{Experimental technique}
The activation measurements were carried out at the Karlsruhe 3.7
MV Van de Graaff accelerator. Neutrons were produced with the
$^7$Li($p,n$)$^7$Be source by bombarding 30 $\mu$m thick layers of
metallic Li on a water-cooled Cu backing with protons of 1912 keV,
30 keV above the reaction threshold. The angle-integrated neutron
spectrum imitates almost perfectly a Maxwell-Boltzmann
distribution for $kT$ = 25.0$\pm$0.5~keV with a maximum neutron
energy of 108~keV \cite{RaK88}. Hence, the proper stellar capture
cross section at $kT$= 25~keV can be directly deduced from our
measurement. For an extrapolation to higher and lower temperatures
we make use of either energy-dependent cross section data, as
available from the libraries JEFF, JENDL and ENDF-B, or normalize
the semi-empirical cross section and energy-dependence of Bao et
al. \cite{bao00} to our experimental result.

The neutron f\-lux is kinematically collimated in a forward cone
with 120$^\circ$ opening angle. Neutron scattering through the Cu
backing is negligible since the transmission is about 98 \% in the
energy range of interest. To ensure homogeneous illumination of
the entire surface, the beam with a DC current of
$\approx$100~$\mu$A was wobbled across the Li target. Thus, the
mean neutron f\-lux over the period of the activations was
$\approx$1.5-2$\times$10$^9$ $n$/s at the position of the samples,
which were placed in close geometry to the Li target. A
$^6$Li-glass monitor at 1~m distance from the neutron target
recorded the time-dependence of the neutron yield in intervals of
1~min as the Li target degrades during the irradiation. This
allows for the later correction of the number of nuclei, which
decayed during the activation. This correction is small for very
long half-lives like the ground state of $^{79}$Se, but becomes
important for the short-lived isomeric state.

Samples of selenium metal and cadmiumselenide (CdSe) with natural
isotopic abundance (23.77\% $^{78}$Se) were used in the
measurements. The sample material was pressed to thin pellets,
which were enclosed in a 15~$\mu$m thick aluminium foil and
sandwiched between 10-30~$\mu$m thick gold foils of the same
diameter. In this way the neutron f\-lux can be determined
relative to the well-known capture cross section of $^{197}$Au
\cite{RaK88}.

For the determination of the partial cross section to the 3.92~min
isomer, we irradiated several thin Se metal samples for 600-710~s,
corresponding to a time-integrated total neutron f\-lux of
$\Phi_{tot}$= 1.2-2.9$\times$10$^{12}$ $n$. The induced
$\gamma$-ray activity of the 95.7~keV transition
(I$_\gamma$=9.62$\pm$0.28\% \cite{NDS79}) was afterwards counted
off-line in a well def\-ined geometry of 76.0$\pm$0.5~mm distance
using a shielded HPGe detector in a low background area. Since we
recorded several 120~s spectra, we were also able to measure the
half-life of the isomer to be 3.90$\pm$0.14~min, in perfect
agreement with the literature value of 3.92$\pm$0.01~min
\cite{NDS79}.

For the AMS measurement one CdSe sample was irradiated for 13d
($\Phi_{tot}$= 1.96$\times$10$^{15}$ $n$). The number of produced
$^{79}$Se nuclei was measured at the Munich 14~MV tandem
accelerator using the gas-f\-illed analyzing magnet system (GAMS)
\cite{KFG00}. The AMS system at Munich consists of a sputter ion
source, a 90$^\circ$ mass analyzer, an 18$^\circ$ electrostatic
def\-lection, a 14 MV tandem accelerator and a Wien f\-ilter. To
separate the radioisotopes from the stable isobars we use the
sensitive combination of a gas-f\-illed magnet with a
multi-$\Delta$E ionization chamber. Since AMS determines the
concentration ratio of a radioisotope to a stable isotope,
[$^{79}$Se]/[$^{78}$Se], relative to a standard of known isotopic
ratio, we can easily deduce our experimental cross section from
the equation $\sigma=\frac{[^{79}Se]}{[^{78}Se]} \times
\frac{1}{\Phi_{tot}}$. The biggest challenge in our case was the
suppression of isobaric contaminations from the stable neighbor
$^{79}$Br (49.31\%), since Bromine is rather volatile and readily
forms negative ions. The contamination of $^{79}$Br is reduced by
means of the gas-f\-illed magnet in front of the ionization
chamber by roughly two orders of magnitude. We chose a rather high
charge state of 15$^+$ (stripping yield: 0.64\%) at a terminal
voltage of 12.55 MV. For a more detailed description of the
$^{79}$Se AMS measurement, refer to \cite{RDF06}.

\section{Preliminary results}
From the activity measurement of the isomeric state, we can deduce
a partial Maxwellian averaged cross section of
$<\sigma>_{30~keV}$(part.)= 42.0$\pm$2.0~mbarn. With AMS we
determined for the total cross section a ratio of
[$^{79}$Se]/[$^{78}$Se]= (1.19$\pm$0.19)$\times$10$^{-10}$,
corresponding to $<\sigma>_{30~keV}$= 60.1$\pm$9.6~mbarn.
Following these results, the isomeric ratio (IR), which should --
according to Hauser-Feshbach theory -- show no energy-dependence,
decreases from 0.88$\pm$0.06 at $kT$=25~meV to 0.70$\pm$0.11 at
$kT$=30~keV.

The rather high uncertainty of the AMS measurement is due to the
high background from $^{79}$Br, and the uncertainty in the used
$^{79}$Se standard. This standard was produced using the thermal
cross sections, which themselves exhibit rather large
uncertainties ($\sigma_{th}$($^{79}$Se$^g$)= 50$\pm$10~mbarn;
$\sigma_{th}$($^{79}$Se$^m$)= 380$\pm$20~mbarn \cite{MDH81}).
Thus, one requirement to reduce the uncertainty is the
(chemistry-free) preparation of an independent standard, e.g. via
the reaction $^{82}$Se$(p,\alpha)$$^{79}$As($\beta^-$,
8.2m)$^{79}$Se. The amount of $^{79}$Se atoms can then be easily
measured with $\gamma$ spectroscopy. With this new standard, a
re-measurement of the thermal and stellar total neutron cross
sections is planned.

\section{Astrophysical implications}
Our result for the total cross section of $^{78}$Se is by a factor
of 2 lower than the previous \mbox{recommended} semi-empirical
cross section from Bao et al. (109$\pm$41~mbarn \cite{bao00}).
Additionally, a recent remeasurement of the total $^{81}$Br cross
section showed the same trend \cite{MH06}, the cross section being
a factor of $\approx$1.4 lower than than the previous recommended
value.

Adding up these two signif\-icantly lower cross sections, we
expect in future stellar model calculations a weaker reaction
f\-lux across the $^{79}$Se branching, which shifts the
$^{80}$Kr/$^{82}$Kr ratio to larger values due to the larger
$^{80}$Kr abundances.

\begin{acknowledgments}
The authors would like to thank E.P. Knaetsch, D. Roller and W.
Seith for their help and support during the irradiations at the
Karlsruhe Van de Graaff accelerator. We gratefully acknow\-ledge
the excellent operation of the Munich MP tandem by the staff
during our experiment. This work was supported by the Swiss
National Science Foundation Grants 2024-067428.01 and 2000-105328,
and by the "Sonderforschungsbereich 375-95 f\"ur
Astro-Teilchenphysik" der Deutschen Forschungsgemeinschaft.
\end{acknowledgments}

\end{document}